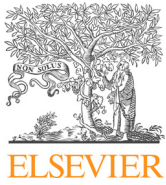
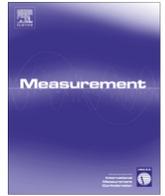

# A recurrent neural network approach for remaining useful life prediction utilizing a novel trend features construction method

Sen Zhao [a], Yong Zhang [b], Shang Wang [a], Beitong Zhou [a], Cheng Cheng [a],⁎

[a] *School of Artificial Intelligence and Automation, Huazhong University of Science and Technology, Wuhan 430074, China*
[b] *School of Information Science and Engineering, Wuhan University of Science and Technology, Wuhan 430081, China*

ABSTRACT

Data-driven methods for remaining useful life (RUL) prediction normally learn features from a fixed window size of a priori of degradation, which may lead to less accurate prediction results on different datasets because of the variance of local features. This paper proposes a method for RUL prediction which depends on a trend feature representing the overall time sequence of degradation. Complete ensemble empirical mode decomposition, followed by a reconstruction procedure, is created to build the trend features. The probability distribution of sensors' measurement learned by conditional neural processes is used to evaluate the trend features. With the best trend feature, a data-driven model using long short-term memory is developed to predict the RUL. To prove the effectiveness of the proposed method, experiments on a benchmark C-MAPSS dataset are carried out and compared with other state-of-the-art methods. Comparison results show that the proposed method achieves the smallest root mean square values in prediction of all RUL.

© 2019 Elsevier Ltd. All rights reserved.

## 1. Introduction

Contrary to the time-based maintenance strategy which checks mechanical systems on a regularly scheduled basis, condition-based maintenance (CBM) reduces the uncertainty of maintenance which making decisions according to the real-time information through condition monitoring. Remaining useful life (RUL) prediction is a key aspect of CBM, which estimates the duration of a component or a system from the current time to the end of its lifetime. An accurate RUL prediction improves the productivity, reliability and safety of a mechanical system while minimizing its maintenance cost. With the technology maturing and costs dropping, large amounts of sensor information have been captured by industry manufacturers and research laboratories, for the purposes of fault and degradation analysis.

Given its importance, RUL prediction has been the subject of many investigations in the field of reliability for modern manufacturing [1,2]. Generally, in these papers, methods for RUL prediction can be classified into model-based and data-driven approaches. Model-based approaches develop mathematical models based on the knowledge of the failure mechanism of the models. However, for some complex systems, such as the engine turbine fan, accurate failure mechanism models are hard to build precisely due to system uncertainties and environment noises.

Data-driven approaches can be divided into statistical model-based approaches and artificial intelligence approaches. Statistical model-based approaches [3] include methods based on the Wiener and Gamma process [4], the Kalman filter method [5–7], and particle filters [8,9]. Statistical models are built according to physical laws or probability distributions. Given an example, in [7], Cui et al. established models for each kind of bearing operation and incorporated unscented Kalman filter into the Bayesian estimation method to calculate the probability of each state. Instead, the artificial intelligence approaches [10–12] build models without knowing the state of the system. They train models based on the historical degradation information by intelligent learning algorithms.

In terms of the intelligent learning algorithms, feature extraction is first performed to capture the key information and also to reduce the dimension of the original signal. Traditional features, such as mean square value in time domain or fast Fourier transform in the frequency domain [13], are often highly noised. In addition, features captured for a mechanical system are commonly not the best representations for another mechanical system. Therefore, a lot of human effort is expended on finding proper features in a case-by-case manner [14–16]. Few of these features consider the uncertainty of the measurement. In fact, industrial time series signals acquired by sensors are intensely disturbed by the unpredictable environment and the uncertainties of the physical systems. Therefore, each measurement point of the time series signal is uncertain itself. However, the tendency of the physical systems during system degradation is relatively certain. In this study, we regard the measurement of each point during degradation as a trend-centered probability distribution. In the Bayesian machine literature, Gaussian Processes (GPs) are popular choices for modeling a distribution over regression functions. Yet the design of the appropriate prior for GPs is hard, especially when modeling the distribution over different sensors' measurements. Conditional Neural Processes (CNPs) [17] are recently developed network systems which solve this problem. Structured as neural networks, CNPs are able to learn appropriate prior from data themselves. Thus, it is natural to use the CNPs to map the sensors' measurements in probability distribution. But as CNPs are still time consuming, the more effective trend features construction is still meaningful. In [18], M. Torres et al. proposed a complete ensemble empirical mode decomposition (CEEMD) method for analyzing transient events, which is an enhancement of the extended empirical mode decomposition (EEMD) [19] and empirical mode decomposition (EMD) [20]. CEEMD decomposes the original signal into a residue and a number of intrinsic mode functions (IMFs), representing low to fast oscillation components of the signal. In [21], Wu, J. et al. used three IMFs to track the degradation evolution of rolling bearings. In [22], M. Niu et al. use three IMFs and the residue of the CEEMD as individual features for four separate SVR models to predict the RUL. In this work, we found that the residual of the CEEMD and the extracted IMFs can also be reconstructed in an optimal way to build general a trend feature for mechanical systems.

When the trend feature of a testing machine is extracted, advanced learning algorithms are commonly followed to train prediction models. The most well-known advanced learning algorithms for RUL prediction modeling include neural networks (NN) [11] and support vector machines (SVM) [23]. SVM is trained to solve a quadratic optimization problem by introducing bound constraints and a linear equality constraint. SVM has been well investigated for RUL prediction, see successful implementations in [24–26]. Recently, deep learning methods (i.e., convolutional neural networks [CNN] and recurrent neural networks [RNN]) have merged into research and industry fields, and yield very accurate results in speech recognition [27] and image recognition [28] that surpass other traditional NN. The methods based on CNN and SVM learn features from a fixed time window size of a priori of degradation, which may lead to less accurate prediction results on a different dataset because of the variance of local features. To improve the prediction performance, the information of the entire degradation process should contribute to constructing a more accurate model. RNN is a method which has been found well-suited for this task due to its ability to remember past input information over an period of time. However, RNN itself suffers a drawback called the 'vanishing error problem' [29] that results in vanishing or exploding gradients when learning long-term dependencies. This drawback raises the problem of which information is valuable to remember and which to forget. [29] proposed the long short-term memory unit (LSTM) RNN which adds memory controllers to the conventional RNN architecture to decide when to select related information for prediction and to discard redundant information over a long period without error vanishing.

In this paper, we apply a signal decomposition technique CEEMD to obtain a number of IMFs and a residue. Several trend features are reconstructed based on the IMFs and the residual. The trend features are evaluated using the probability distribution of sensors' measurement learned by CNPs. The best trend feature is chosen and used as the input feature for a model developed using LSTM (denoted as DLSTM) to predict the RUL. The learning capabilities and prediction accuracy of the proposed DLSTM RUL prediction method, with respect to other state-of-the-art approaches, are evaluated through extensive experiments using a benchmark turbine engine dataset. The proposed trend features with the DLSTM prediction method are characterized by the following main features:

1. Proposing a novel trend feature construction method by using the CEEMD method.
2. Proposing a new perspective on the measurement of the sensors, as well as conducting experiments that evaluate several trend features using the distribution of the measurement learned by the neural processes.
3. LSTM RNN, using the novel trend features as inputs, performing more precise RUL predictions when compared to the Kalman filter, SVR, and CNN methods.

The rest of the paper is organized as follows: The methodology of developing the overall RUL prediction framework is explained in Section 2. Section 3 presents the experimental studies of RUL prediction on a turbofan engine dataset and provides the evaluation and comparison of our proposed RUL method, with respect to other existing research results on the same dataset. Finally, Section 4 concludes this paper.

## 2. Methodology

The central idea of this project is illustrated in Fig. 1, which consists of a trend feature extraction process described in Section 2.1 and a LSTM modeling process described in Section 2.2.

### 2.1. Trend feature extraction

#### 2.1.1. Ensemble empirical mode decomposition

Given a dataset $D = [x(1), \ldots, x(n)] \in \mathbb{R}^n$, which can be decomposed into several intrinsic mode functions (IMFs) (or modes) by empirical mode decomposition (EMD). The IMFs have to possess the following two properties: 1) Over the entire time range, the function must have the same number of local extreme points and zero crossing, or only have difference by one; and 2) The envelope of the local maximum (upper envelope) and local minimum (lower envelope) must be zero on average.

As is defined by EEMD [19], the k-th 'true' IMF (denoted $\overline{\text{IMF}}_k$) means the average of IMFs obtained by executing EMD on an ensemble of traits generated by adding different uniformly distributed Gaussian white noises to the original dataset *D*. The $\overline{\text{IMF}}_k$ can be calculated through the following expressions:

$$\bar{x}^i(t) = x(t) + w^i(t), \quad t \in 1, \cdots, n, \tag{1}$$

$$\mathbf{IMF}^i(t) = \text{EMD}(\bar{x}^i(t)) \tag{2}$$

where $\mathbf{IMF}^i(t) = \left[\text{IMF}^i_1(t), \ldots, \text{IMF}^i_k(t)\right]$, then

$$\overline{\text{IMF}}_k(t) = \frac{1}{I}\sum_{i=1}^{I} \text{IMF}^i_k(t) \tag{3}$$

where $w^i(t) \in \mathbb{R}^n$ denotes the i-th realization of the white noises for $i = 1, \cdots, I$. EMD (·) represents the original EMD function proposed in [20], which could generate *k* number of IMFs for i-th realization.

#### 2.1.2. Complete ensemble empirical mode decomposition

CEEMD [18] works better than EEMD with lower computing cost and smaller reconstruction error. It realizes these advantages by adding Guassian noises (for $i = 1, \cdots, I$) on each residual $r_k$.

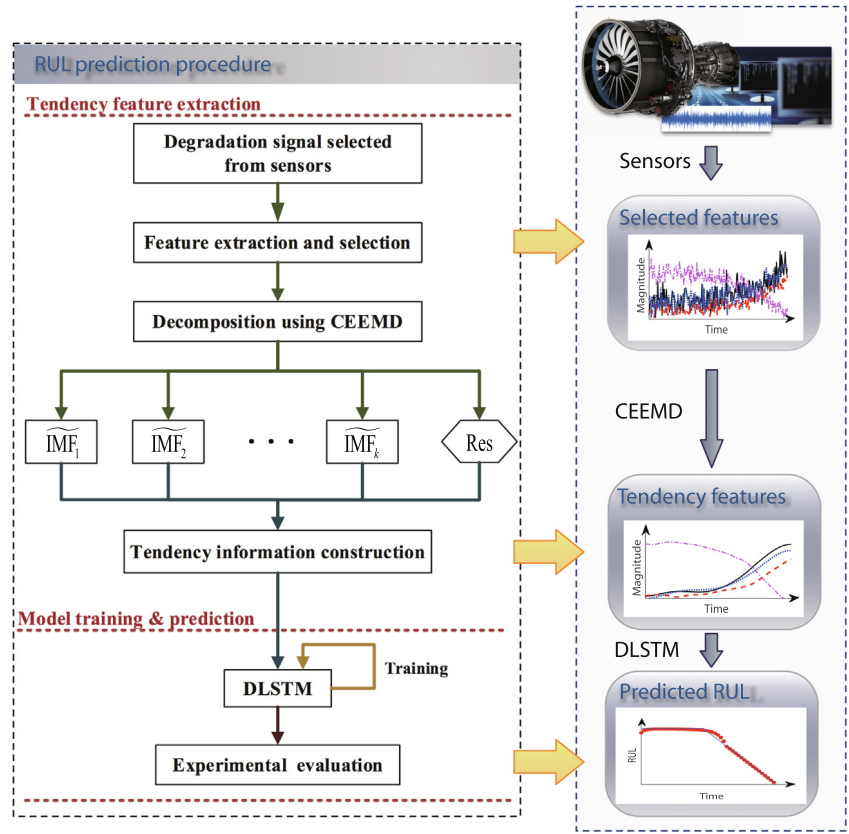

**Fig. 1.** The overview of the proposed method.

$\widetilde{IMF}_k(t)$ denotes the k-th mode of CEEMD. Now compute the $\widetilde{IMF}_k(t)$ and the first residual $r_1$:

$$\bar{x}^i(t) = x(t) + \varepsilon_0 w^i(t), \ t \in 1, \cdots, n \tag{4}$$

$$IMF_1^i(t) = EMD_1(\bar{x}^i(t)) \tag{5}$$

$$\widetilde{IMF}_1(t) = \frac{1}{I}\sum_{i=1}^{I} IMF_1^i(t) \tag{6}$$

$$r_1(t) = x(t) - \widetilde{IMF}_1(t) \tag{7}$$

Eqs. (4)–(7) are defined as the first step of CEEMD. It can be viewed that $\widetilde{IMF}_1(t)$ is the first mode of EEMD. $x(t)$ denotes the original signal, $w^i(t)$ denotes the i-th realization of $\mathcal{N}(0,1)$, $\varepsilon_k$ denotes the noise standard deviation (NSD) for k-th residual and $\varepsilon_0$ denotes the NSD for the original signal $x(t)$. In Eq. (5), $EMD_1(\cdot)$ represents the first step of the EMD in which the first IMF mode is created.

Then, define the second step of CEEMD by the following equations to calculate k-th residual, $r_k$, and k-th IMF, $\widetilde{IMF}_k(t)$:

$$\bar{r}_k^i(t) = r_k(t) + \varepsilon_k w^i(t) \tag{8}$$

$$\widetilde{IMF}_k(t) = \frac{1}{I}\sum_{i=1}^{I} EMD_1(\bar{r}_{k-1}(t)) \tag{9}$$

$$r_k(t) = r_{k-1}(t) - \widetilde{IMF}_{k-1}(t) \tag{10}$$

Eqs. (8)–(10) are repeated until the residual $\bar{r}_k(t)$ can not be decomposed any more (i.e., $\bar{r}_k(t)$ does not have at least two extremes). The third step is to obtain the final residual by

$$R(t) = x(t) - \sum_{k=1}^{K} \widetilde{IMF}_k(t) \tag{11}$$

This means that the original signal can be reconstructed by adding up the final residual and $\widetilde{IMF}_k(t)$ for $k = 1, \ldots, K$

$$x(t) = R(t) + \sum_{k=1}^{K} \widetilde{IMF}_k(t). \tag{12}$$

### 2.1.3. Trend feature construction

As has been mentioned that the $\widetilde{IMF}_k(t)$ is the fluctuation signal of $x(t)$, $\widetilde{IMF}_k(t)$ with smaller $k$ contains more information with higher frequency oscillation but also more noise. In order to balance the containment of information and noise, the trend feature signal $Q(t)$ is finally reconstructed,

$$Q_v(t) = \begin{cases} R(t) + \sum_{k=m}^{K} \widetilde{IMF}_k(t), & \text{for } m \leqslant K \\ R(t), & \text{for } m = K+1 \end{cases} \tag{13}$$

where $v = K + 1 - m$, and $m \in \{1, 2, \ldots, K+1\}$ controls the usage of the $\widetilde{IMF}_k(t)$.

### 2.2. Model training and predicting

Using the constructed trend features as the inputs, in this section, a DLSTM framework for RUL prediction is proposed. Section 2.2.1 describes how the LSTM network makes predictions over a long time without the impact of local noises and the problem of error vanishing. Section 2.2.1 and 2.2.3 explain two techniques that help to train the network. Section 2.2.4 summarizes the DLSTM algorithm.

### 2.2.1. Concept of Vanilla LSTM

Vanilla LSTM is the most widely used LSTM architecture variation. The architecture of the vanilla LSTM is shown in Fig. 2.

To summarize, the core idea of LSTM lies in two information flows and includes three gates, which function is to add and remove information. More specifically, two information flows are detailed as illustrated in Fig. 2: The lower green flow controls the combination of the short-term state of the last cell ($h_{t-1}$) and the input of the current cell ($Q_t$). During this process, three gates named forget gate, input gate, and output gate are used to generate three signals named forget control signal, input control signal, and the output control signal (denoted $f_t$, $i_t$, and $o_t$ respectively); the upper brown flow shows the update of the current cell's long-term state ($C_t$). Two branches are introduced into the upper flow, including the long-term state of the last cell ($C_{t-1}$) controlled by $f_t$ and the current input value controlled by $i_t$. More details of LSTM are explained below:

In terms of the forget gate, the state information ($f_t$) is captured from the short-term state of last cell ($h_{t-1}$) and the current input ($Q_t$), and goes through a sigmoid function $\sigma(\cdot)$:

$$f_t = \sigma(W_f \cdot Q_t + R_f \cdot h_{t-1} + b_f), \qquad (14)$$

where $W_f, R_f$ are the weights, and $b_f$ is the bias parameter.

For the input gate, this gate generates the input information ($i_t$), which is captured from the short-term state of last cell ($h_{t-1}$) and the current input ($Q_t$), and goes through a sigmoid function to obtain $i_t$:

$$i_t = \sigma(W_i \cdot Q_t + R_i \cdot h_{t-1} + b_i), \qquad (15)$$

where $W_i, R_i$, and $b_f$ are the weights and bias parameters.

Then, the long-term state of current cell ($C_t$) is updated by the forget control signal ($f_t$) and input control signal ($i_t$) obtained in Eq. (14) and Eq. (15):

$$C_t = f_t \ast C_{t-1} + i_t \ast \tilde{C}_t \qquad (16)$$

where $\tilde{C}_t = tanh(W_c \cdot Q_t + R_c \cdot h_{t-1} + b_c)$.

Here $\tilde{C}_t$ is the candidate memory cell gate, $W_c$ and $R_c$ are the weights for $Q_t$ and $h_{t-1}$, and $b_c$ is the bias value.

The bottom data flow goes through the output gate where the output control signal $o_t$ is generated by the following equation:

$$o_t = \sigma(W_o \cdot Q_t + R_o \cdot h_{t-1} + b_o) \qquad (17)$$

where $W_o$ and $R_o$ are the weights for $h_{t-1}$ and $Q_t$, and $b_o$ is the bias value.

The output of the current cell ($Z_t$) and the short-term cell state of the current cell ($h_t$) is generated by multiplying the current long-term cell state ($C_t$) by the output control signal ($o_t$) and it is also transferred to the next cell through the bottom data flow,

$$Z_t = h_t = o_t \ast \tanh(C_t). \qquad (18)$$

### 2.2.2. Concept of dropout

G. Hinton et al. proposed dropout in [30] which aims to solve the problem of overfitting. Each neuron in the dropout layer is randomly omitted from the network with a probability $P_c$ within (0,1) while training. Networks vary through training cases, but all networks share the same weights. All neurons are activated in testing cases, where the weights are calculated by multiplying the training weight and $P_c$.

### 2.2.3. Masking layer

Generally, the proposed model should fit and make predictions on different samples with different lengths, however, the length of the RNN is a pre-defined value. A masking layer is introduced to tackle this problem. Samples are first extended to the longest sample length by padding in front with zeros. The masking layer will work by skipping the points where all the input features are zeros.

### 2.2.4. DLSTM

According to the above descriptions, the proposed DLSTM framework (see Fig. 3) for RUL prediction can be summarized into Algorithm 1.

**Algorithm 1** Outline of DLSTM training for RUL estimation [31].

**Input** : The extracted label **RUL** $\in \mathbb{R}^{1 \times n}$;
The trend features $Q_v \in \mathbb{R}^{N_J \times n}$, where $N_J$ is the number of time-series measurement.
**Output:** Trained neural networks
**Initialize:** LSTM layer parameters and dense layer parameters.
**repeat**
    **Forward Propagation:**
    Using masking layer to skip the zero points.
    **do**
        Conducting LSTM operation with the trend sequence using Eq. (14)~(18).
    **end**;
    Dropout layer is employed to avoid the overfitting problem with the probability $P_c = 0.5$.
    Conventional fully-connected network is used for RUL regression.
    The *tanh* function is introduced for normalized output.
    Compute the loss with the loss function MSE.
    **Backward Propagation:**
    Compute the gradient using *Adam* [31] and update network parameters.
**until** *Maximum iterations*;
Use the trained DLSTM to estimate the RUL on the test engine units.

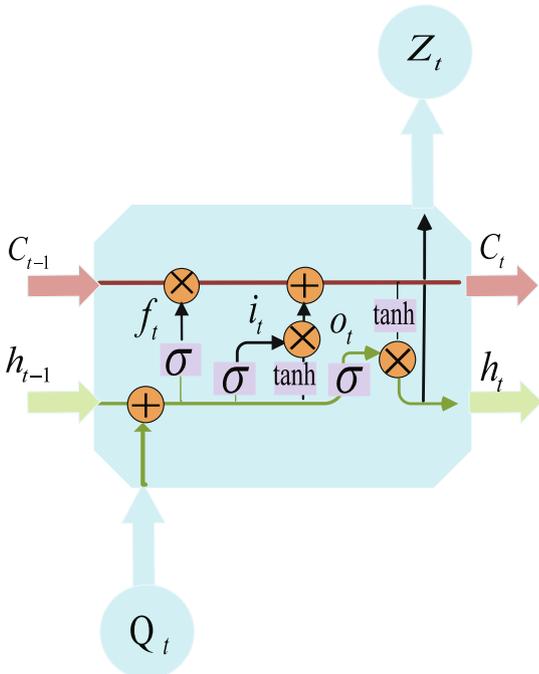

**Fig. 2.** The vanilla architecture of the LSTM RNN predictor.

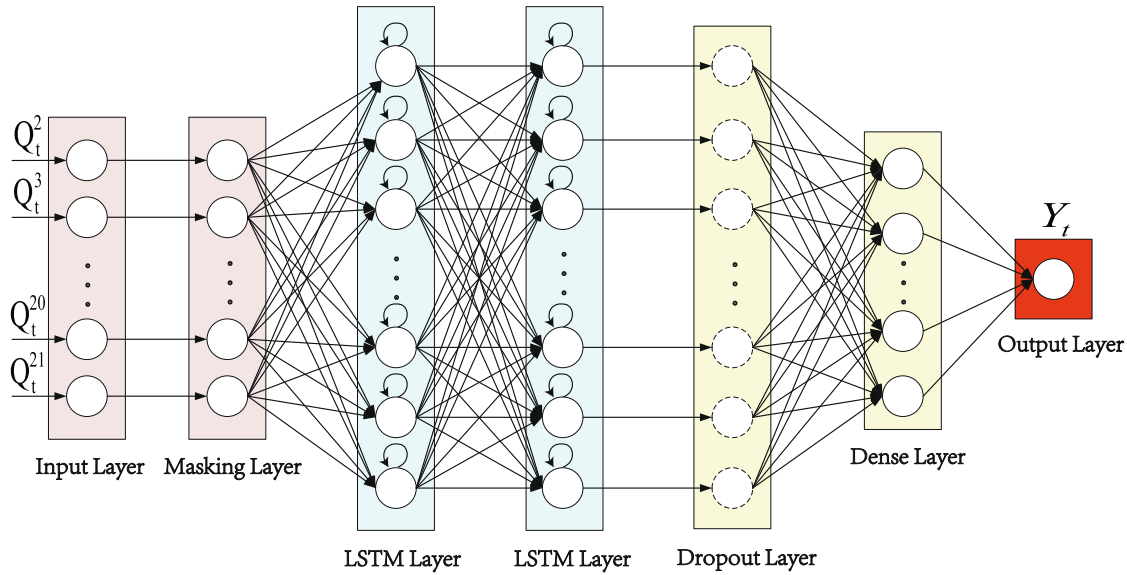

**Fig. 3.** The configuration of the DLSTM. The DLSTM contains one masking layer, two RNN layers, one dropout layer, and one dense layer.

## 3. Experimental study

### 3.1. Data description

In this paper, NASA's turbofan engine RUL prediction problem [32] is employed to evaluate the proposed method, which was produced on the commercial modular aero-propulsion system simulation (C-MAPSS) test-bed. The detailed information of the C-MAPSS dataset are presented in the Table 1. The dataset contains four subsets, named as FD001, FD002, FD003, and FD004. Engine units for each subset for training and testing are listed in the first and second row of Table 1 respectively. FD001 and FD002 are with fault mode 1, while FD003 and FD004 are with fault mode 2.

The time-series measurement signals used for training and prediction are acquired from 21 sensors under three operational conditions associated with different sea levels. The operational status of each engine unit is healthy in the early stage and begins to degrade over time until a failure occurs. The predicted RUL indicates the time periods over which those engines are likely to operate before they require repair or replacement.

The training datasets contain the whole evolution of the engines' degradation until the end of their lifetimes, which means the RUL value at the last time point is 0. In testing datasets, only the first partial degradation is given of each engine, hence, the RUL at the last point is not 0. The objective of this study is to precisely predict the RUL during the system degradation till the last point on the testing datasets. In this project, models are trained individually based on different subsets. For each modeling training process, the training data is used as the input and the corresponding RUL is used as the label. Then the trained models are applied to the test datasets for prediction and evaluation.

Programming languages are Python 3.6 with deep learning library Keras, and Matlab 2018a. Both programs are accelerated by GPU.

**Table 1**
Details of the C-MAPSS dataset.

|  | FD001 | FD002 | FD003 | FD004 |
|---|---|---|---|---|
| Engine units, training | 100 | 260 | 100 | 249 |
| Engine units, testing | 100 | 259 | 100 | 248 |
| Operation conditions | 1 | 6 | 1 | 6 |
| Fault modes | 1 | 1 | 2 | 2 |

### 3.2. Data pre-processing

C-MAPSS dataset contains time-series measurements from 21 sensors. However, measurements of some sensors hardly change over time, indicating there are no useful information in these measurements. We abandon these measurements and keep the remaining ones. Finally, the available sensor datasets are $D^J = \left[D^2, D^3, D^4, D^7, D^8, D^9, D^{11}, D^{12}, D^{13}, D^{15}, D^{17}, D^{20}, D^{21}\right]$ with sensor index $j \in J$, $J = \{2, 3, 4, 7, 8, 9, 11, 12, 13, 15, 17, 20, 21\}$.

Before feature extraction, the measurement data are first normalized by using the min-max normalization method. $D_{norm}$ denotes the normalized dataset. $D_{norm} = [x_{norm}(1), \ldots, x_{norm}(n)] \in \mathbb{R}^n$. The normalized measurement data $x_{norm}^j(t)$ are computed by:

$$x_{norm}^j(t) = \frac{2\left(x^j(t) - x_{min}^j\right)}{\left(x_{max}^j - x_{min}^j\right)} - 1, \quad \forall t, j \quad (19)$$

where $x^j(t)$ is the t-th measurement of the j-th sensor. $x_{max}^j$ and $x_{min}^j$ denote the maximum and minimum values of measurements of the j-th sensor, respectively.

*Training set augmentation*: Sequences in the training sets are running to the end of engines' lifetime (i.e., RULs equivalent to 0), however, sequences in the testing sets always end with nonzero RULs. To solve this problem, one step of augmentation is proposed. In this step, the last 1 to 100 points are cropped out from the end of different sequence in the training set, thus, we get some training sequences end with the RUL from 1 to 100.

*Set the RUL for the training set*: Considering the fact that the engine units work in its good health condition in a very long time period, the degradation in this period is negligible to estimate. In this work, we follow the suggestion in [33], in which indicates that the faults usually occur at the time step around 120~130. We set the RUL bigger than 130 to be 130.

### 3.3. Performance metrics

In this study two metrics are used to evaluate the performance of the proposed prognostic method, i.e. score function and root mean square error (RMSE).

The score function is defined by 2008 PHM Data Challenge Competition [34]. The score function is given as follows:

$$s = \sum_{p=1}^{P} s_p, s_p = \begin{cases} e^{-\frac{d_p}{13}} - 1, & \text{for } d_p < 0 \\ e^{\frac{d_p}{10}} - 1, & \text{for } d_p \geq 0 \end{cases} \quad (20)$$

where $d_p = \widehat{RUL}_p - RUL_p$, $\widehat{RUL}_p$ and $RUL_p$ denote the predicted RUL and the label of p-th engine unit, respectively.

RMSE for each sub-set is obtained by the following equation:

$$\text{RMSE} = \sqrt{\frac{1}{P} \sum_{p=1}^{P} d_p^2} \quad (21)$$

### 3.4. Trend feature construction using CEEMD

CEEMD is used to obtain clear trend features for measurement data of each engine unit. For CEEMD implementation, the noise standard deviation is set to 0.02, the number of realization is set to be 100, and the maximum iterations is set to 5000.

To choose the optimal selection of $m$, three trend features $Q_0^j, Q_1^j$, and $Q_2^j$ for $j = 2, 8, 12, 15$ of engine unit 1 in FD001 are reconstructed by substituting $m = K+1, K, K-1$ into Eq. (13), respectively. Fig. 4 shows the evolutions of $Q_0^j, Q_1^j, Q_2^j$ and $D_{norm}^j$ of engine unit 1 in FD001 sub-set. It is observed that the $Q_0^j$ trend feature in Fig. 4(a) with $m = K+1$ offers the most clear trend (with less fluctuation) compared with $x_{norm}^j$ in Fig. 4(d), $Q_1^j$ in Fig. 4(b), and $Q_2^j$ in Fig. 4(c). The trend features will be further evaluated in subSection 3.5.

### 3.5. Trend features evaluation

CNPs are employed to learn the probability density for the measurement of the sensors. By setting the probability distribution as the baseline for tendency features, we numerically compare $Q_0, Q_1, Q_2, D_{norm}$ with other three tendency features constructed by EEMD (i.e., $E_0, E_1$, and $E_2$), one popular time-domain statistical features (mean). $E_0, E_1$, and $E_2$ are constructed through the following expression:

$$E_v(t) = \begin{cases} R'(t) + \sum_{k=m}^{K} \overline{IMF}_k(t), & \text{for } m \leq K \\ R'(t), & \text{for } m = K+1 \end{cases} \quad (22)$$

where $R'(t)$ and $\overline{IMF_k(t)}$ denote the residual and the intrinsic mode functions extracted based on $x_{norm}(t)$ using EEMD method (more technique details of EEMD can refer to [19]).

mean value is computed through:

$$mean(t) = \frac{1}{T} \sum_{n=n_1}^{n_2} x_{norm}(n) \quad (23)$$

where $n_1 = t - (T-1)/2, n_2 = t + (T-1)/2$, T = 5.

Two evaluation metrics including the mean probability density (MPD) and the mean absolute relative error (MARE) are employ to assess these features:

$$\text{MPD} = \frac{1}{C} \sum_c \frac{1}{T_c} \sum_t N(x_{c,t} | \mu_{c,t}, \sigma_{c,t}). \quad (24)$$

$$\text{MARE} = \frac{1}{C} \sum_c \frac{1}{T_c} \sum_t |x_{c,t} - \mu_{c,t}|. \quad (25)$$

where $x_{c,t}$ is the trend feature for engine unit c and time period t, $\mu_{c,t}$ is the median for the probability distribution. $\sigma_{c,t}$ is the standard derivative for the probability distribution. C is the number of engine units, $T_c$ is the number of time periods for engine unit c. c = 1, 6, 7, 8, 12, 19, 26, 29, 38, 39,42, 49, 50, 58, 60, 62, 63, 68, 73, 78, 81, 86, 91, 94, 98. These engine units in FD001 training set are randomly chosen to evaluate the features.

Comparison results are summarized in Tables 2, 3, Figs. 6, 7. It can be observed that $Q_0$ obtains the highest MPD and the lowest MARE over all sensors, the lowest RMSE and Score over all datasets, demonstrating that $Q_0$ is the best choice of representing the degradation tendency of this mechanical system.

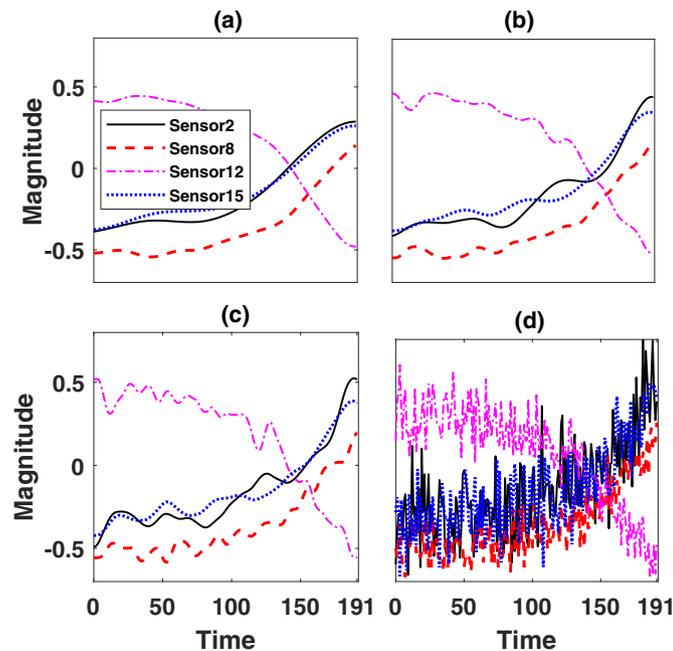

**Fig. 4.** Trend features (a) $Q_0^j$, (b) $Q_1^j$, (c) $Q_2^j$, and (d) $D_{norm}^j$ of engine unit 1 in FD001 sub-set. We plot the features with sensor number $j = 2, 8, 12, 15$.

**Table 2**
Feature evaluation using MPD.

| Features | C-MAPSS FD001 | | | | | | |
|---|---|---|---|---|---|---|---|
| | Sensor 2 | Sensor4 | Sensor 7 | Sensor 13 | Sensor 15 | Sensor 20 | Sensor 21 |
| **$Q_0$** | **2.4263** | **2.9676** | **2.9463** | **3.5390** | **2.7437** | **2.6644** | **2.6176** |
| $Q_1$ | 2.4031 | 2.9354 | 2.9076 | 3.4994 | 2.7157 | 2.6225 | 2.5836 |
| $Q_2$ | 2.3782 | 2.8948 | 2.8794 | 3.4614 | 2.6894 | 2.5809 | 2.5555 |
| $E_0$ | 2.0125 | 2.2961 | 2.2808 | 2.0737 | 2.0581 | 2.3325 | 2.1977 |
| $E_1$ | 2.3465 | 2.8575 | 2.9167 | 3.3679 | 2.7262 | 2.6428 | 2.5783 |
| $E_2$ | 2.2600 | 2.7879 | 2.8472 | 3.3026 | 2.6414 | 2.5546 | 2.5055 |
| mean | 2.2499 | 2.7926 | 2.7733 | 3.3562 | 2.5487 | 2.4534 | 2.4461 |
| $D_{norm}$ | 1.6707 | 2.1422 | 2.1393 | 2.7745 | 1.9055 | 1.8368 | 1.8433 |

**Table 3**
Feature evaluation using MAE.

| Features | C-MAPSS FD001 | | | | | | |
|---|---|---|---|---|---|---|---|
| | Sensor 2 | Sensor4 | Sensor 7 | Sensor 13 | Sensor 15 | Sensor 20 | Sensor 21 |
| **$Q_0$** | **0.0353** | **0.0301** | **0.0307** | **0.0255** | **0.0310** | **0.0330** | **0.0357** |
| $Q_1$ | 0.0423 | 0.0353 | 0.0365 | 0.0293 | 0.0371 | 0.0394 | 0.0413 |
| $Q_2$ | 0.0485 | 0.0407 | 0.0413 | 0.0326 | 0.0422 | 0.0454 | 0.0458 |
| $E_0$ | 0.0935 | 0.1032 | 0.0951 | 0.1443 | 0.1053 | 0.07628 | 0.0880 |
| $E_1$ | 0.0479 | 0.0436 | 0.0348 | 0.0371 | 0.0346 | 0.0375 | 0.0396 |
| $E_2$ | 0.0639 | 0.0521 | 0.0443 | 0.0429 | 0.0476 | 0.0502 | 0.0516 |
| mean | 0.0691 | 0.0514 | 0.0534 | 0.0407 | 0.0587 | 0.0626 | 0.0622 |
| $D_{norm}$ | 0.1449 | 0.1085 | 0.1087 | 0.0768 | 0.1255 | 0.1279 | 0.1287 |

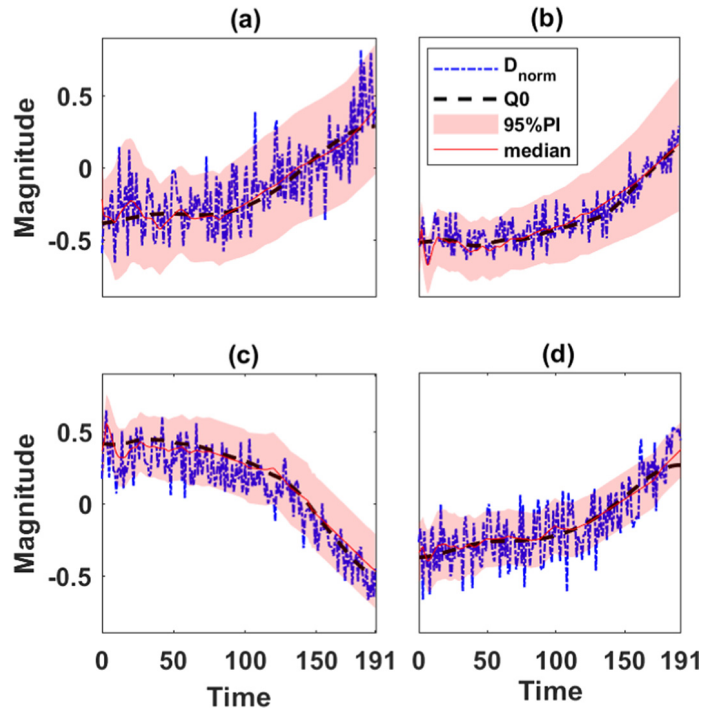

**Fig. 5.** Probability distribution, trend feature $Q_0$ and $D_{norm}$ for sensor 2 (a), sensor 8 (b), sensor 12 (c), sensor 15 (d) of engine unit 1 in FD001 sub-set.

The temporal evolution of tendency feature $Q_0$, normalized original signal $D_{norm}$, 95% probability interval (PI) distribution, and the median curve of the distribution from Sensor 2, 8, 12, and 15 are also shown in Fig. 5. It is observed that $Q_0$ is almost overlapped with the median curve of the distribution. We can conclude that the measurements along the $Q_0$ are very close to the highest probability of measurement points for system degradation.

### 3.6. DLSTM modeling and optimization

To optimize the network architecture, we train and test several models with different architectures on subset FD001. The most suitable architecture for this task is comprised of one masking layer, two LSTM layers, one dropout layer, and one fully-connected layer. Moreover, RUL prediction performance is also varied by changing the output number of the LSTM layer. Table 4 presents the score and RMSE of several configurations. It can be seen that the configuration with the first LSTM layer of 128 outputs and the second LSTM layer with 100 outputs show better prediction performance than other configurations. Therefore, experiments in this project use this configuration.

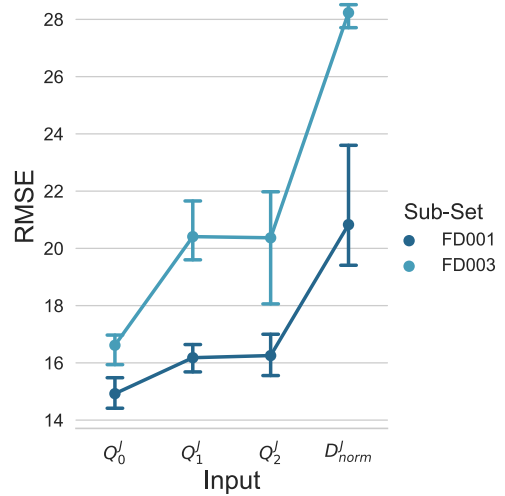

**Fig. 6.** RMSEs of DLSTM model with $Q_0^J, Q_1^J, Q_2^J$, or $D_{norm}^J$ trend feature, on FD001 and FD003 test sets.

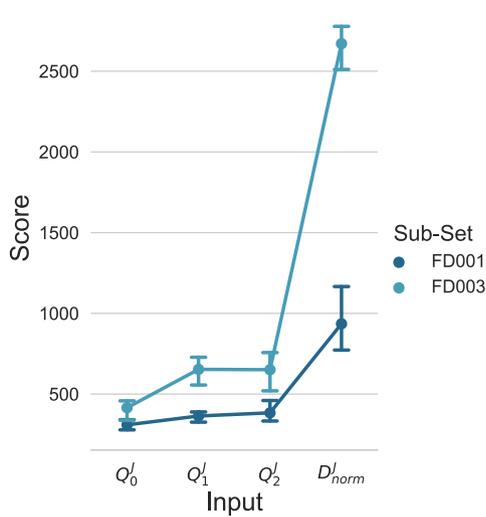

**Fig. 7.** Scores of DLSTM model with $Q_0^J, Q_1^J, Q_2^J$, or $D_{norm}^J$ trend feature, on FD001 and FD003 test sets.

**Table 4**
Configuration evaluation on FD001 test set.

| Output number | | Score | RMSE |
|---|---|---|---|
| 1st LSTM Layer | 2nd LSTM Layer | | |
| 156 | 128 | 819 | 19.59 |
| 128 | 100 | **262** | **14.72** |
| 128 | 64 | 650 | 18.27 |
| 128 | 32 | 380 | 15.26 |
| 64 | 32 | 724 | 18.86 |
| 32 | 16 | 814 | 20.68 |

### 3.7. Prognostic results

Fig. 8 shows RUL prediction results of engine unit numbers 5, 33, 43, and 77 which have been chosen at random from 100 engine units in the FD001 subset. In Fig. 8, during the healthy condition of the engine unit (flatten stage of the label), the proposed method shows precise RUL prediction where the red lines are almost overlapping with the blue lines. When degradation begins, the prediction translates smoothly and then decreases almost linearly with time until the end of the period. All of the four engine units show precise and accurate RUL predictions when close to the end of the period. Near the point of degradation, it appears that the evolution of the predictions does not match the given labels. This is due to the fact that not all of the engine units begin to degrade when RUL equals to 130 and not all of them degrade suddenly without transition. Meanwhile, for the implementation of the proposed method in industry, the accuracy of the prediction is of more importance when the engines work well and when they go to fault. The accurate predictions can reduce the maintenance cost as well as guarantee the safety of the system.

To further investigate the prediction of all the engine units in one subset, we summarize the last RUL prediction values of all the 100 engine units in FD001 in Fig. 9, where the last point values are sorted by ascending number. It can be seen that the prediction values is closer to the label values when the engine units are close to the end of their lives (i.e., RUL < 50). This result indicates that the less the RUL there is, the higher of the prediction accuracy achieved. This is because when the engine unit is about to break down, the trend features of the degradation are enhanced and are captured more easily by the DLSTM.

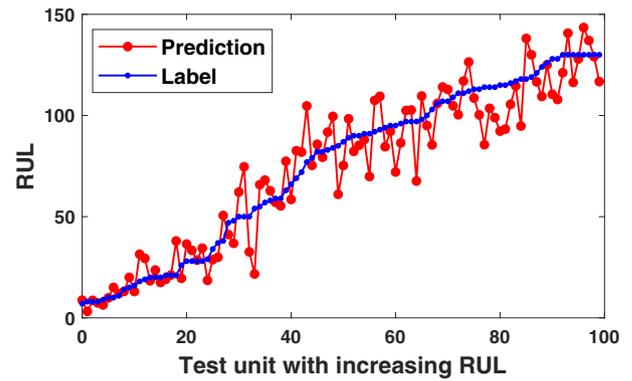

**Fig. 9.** The last RUL prediction values of all the 100 engine units in subset FD001, which are sorted by ascending number.

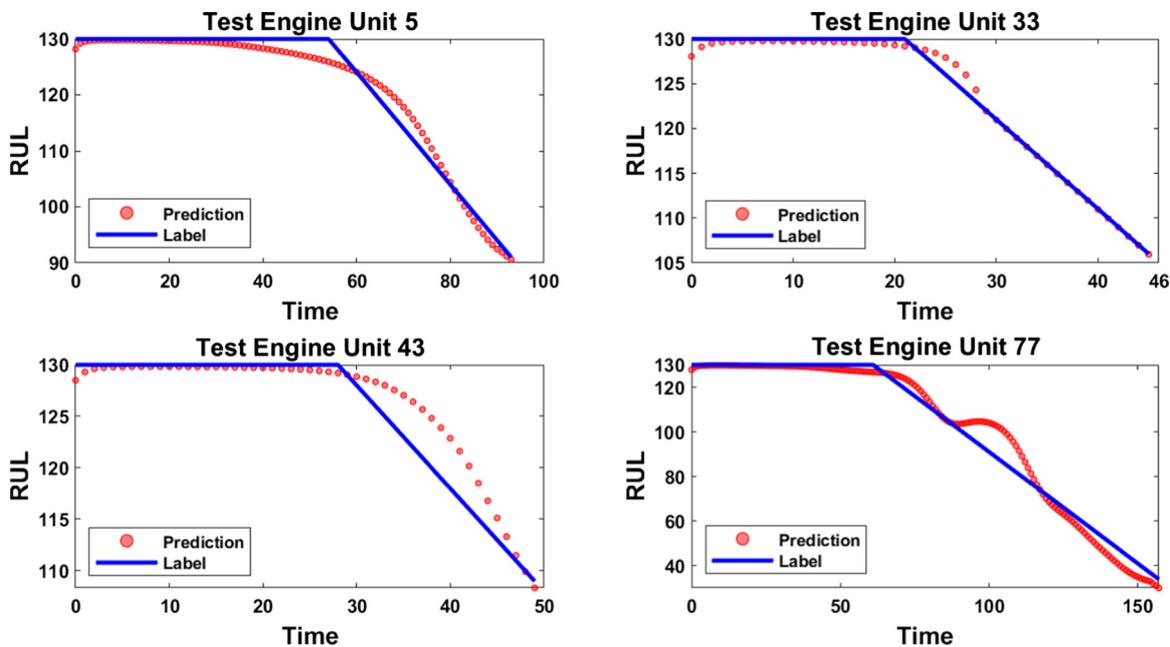

**Fig. 8.** RUL prediction results of engine unit 5, 33, 43, and 77 in FD001 test set.

### 3.8. Comparison

The performances of the different methods evaluated on the FD001 test set are presented in Table 5. The proposed method (DLSTM) is compared with the following methods:

1. RNN with Gate Recurrent Unit (GRU): Besides the difference that the GRU does not use a memory unit to control the flow of the information like the Vanilla LSTM, the RNN with GRU has the same architecture as the DLSTM in Fig. 3.
2. Standard RNN: The standard RNN has one masking layer, three RNN layers, one dropout layer, and one dense layer. Compared to the DLSTM, the standard RNN has one more RNN layer with 50 units added before the dropout layer.
3. CNN: The CNN has three convolution layers, one dropout layer, and two dense layers.

The DLSTM yields a much better performance than the standard RNN, RNN with GRU and CNN with respect to both Score and RMSE results, and achieves 262 on Score and 14.72 on RMSE, respectively. Without trend feature extraction, standard RNN can also work well on the FD001 test set, but needs to contain more layers. The CNN method can also give accurate predictions when the engine units are about to break down, but it does not have the prediction as continuous as the DLSTM.

In Fig. 10 we show the comparison of the RUL predictions of DLSTM and CNN on the 58th engine unit in the FD001 test set. Since the CNN method is based on a time window with the length of 25 [35], the prediction of CNN starts at the time of 24. The DLSTM prediction has a more continuous prediction (a smoother curve) than CNN. Meanwhile, the DLSTM has a better prediction when the engine unit is still in healthy condition. These performances of DLSTM benefit from considering of the overall trend as a learning feature. With the overall trend, the prediction of the RUL will not be influenced by the noises on local features.

Table 6 compares the method proposed in this paper with several existing methods including switching Kalman filter ensemble [36], SVR [15], deep CNN [38], and vanilla LSTM [37]. From Table 6, last row, we see that the proposed method shows better performance with its RMSE and Score metrics are both having smaller values for all the four subsets in C-MAPSS dataset. More specifically, compared with SVR and CNN where predictions are based on the local features, the proposed method considers the overall trend feature of the sequence and results in a smoother degradation curve; Compared with the model-based method such as the Kalman filter, the proposed method offers more flexibility in adjusting the architecture and has better performance in learning the hidden relationship between features and the prediction; Furthermore, the trend feature extracted by CEEMD in this project helps to prevent the problem of the overfitting of vanilla LSTM which uses noised feature to train the model.

**Table 5**
Method evaluation on the FD001 test set.

| Method | Score | RMSE |
|---|---|---|
| RNN with GRU | 745.21 | 18.31 |
| CNN | 491.78 | 18.91 |
| Standard RNN | 367.31 | 15.07 |
| **DLSTM** | **262** | **14.72** |

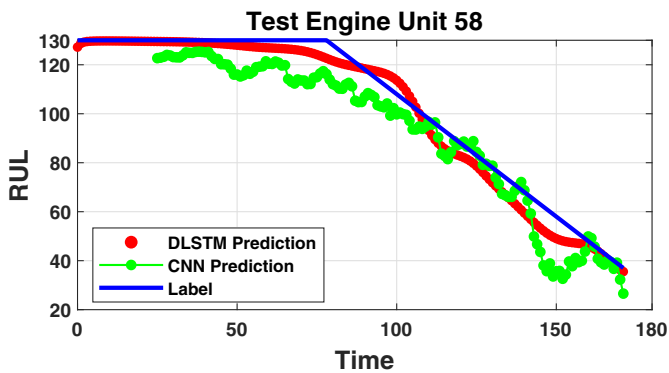

**Fig. 10.** Comparison of the RUL prediction performance between the proposed DLSTM method and the CNN model, on the 58th engine unit in the FD001 test set.

### 4. Conclusion

In this paper, we proposed a novel RUL prediction method by using complete ensemble empirical mode decomposition (CEEMD) for trend feature construction and long short-term memory recurrent neural network (LSTM RNN) for prediction, which are able to prevent the impact of signal fluctuation as well as minimize the prediction error. The trend feature construction associated with mechanical system degradation, which is the major challenge in this study, is constructed according to the combination of intrinsic mode functions (IMFs) and the signal residue. The proposed RUL prediction method is examined on a benchmark dataset of aircraft turbofan engines, and compared to the Kalman filter, support vector regression, deep CNN, and vanilla LSTM. Results verified that the proposed method achieved smaller prediction errors.

As a future project, the adoption of the proposed method in more complex mechanical systems will be considered. Furthermore, we aim to find what hidden information can be extracted from different oscillations to further improve RUL prediction accuracy.

**Table 6**
Comparison between different methods.

| Methods | C-MAPSS | | | | | | | |
|---|---|---|---|---|---|---|---|---|
| | FD001 | | FD002 | | FD003 | | FD004 | |
| | RMSE | Score | RMSE | Score | RMSE | Score | RMSE | Score |
| Switching Kalman Filter Ensemble [36] | [18.4,18,9] | [560,580] | [30,34] | – | [20,22] | [567,610] | [34,37] | – |
| Support Vector Regression [15] | – | 448.7 | – | – | – | – | – | – |
| Deep CNN [37] | 18.45 | – | 30.29 | – | 19.82 | – | 29.16 | – |
| Vanilla LSTM [38] | 19.74 | – | – | – | 24.04 | – | – | – |
| **Proposed CEEMD + DLSTM** | **14.72** | **262** | **29** | **6953** | **17.72** | **452** | **33.43** | **15069** |

**Acknowledgment**

This work was supported in part by the National Natural Science Foundation of China [Grant No. 61873197], and in part by the Primary Research and Development Plan of Jiangsu Province [Grant No. BE2017002].

**References**

[1] B.D. Youn, C. Hu, P. Wang, Resilience-driven system design of complex engineered systems, J. Mech. Des. 133 (10) (2011) 101011.
[2] Y. Zhang, Z. Wang, F.E. Alsaadi, Detection of intermittent faults for nonuniformly sampled multirate systems with dynamic quantization and missing measurements, Int. J. Control (2018) 1–23 (just-accepted).
[3] X.-S. Si, W. Wang, C.-H. Hu, D.-H. Zhou, Remaining useful life estimation–a review on the statistical data driven approaches, Eur. J. Oper. Res. 213 (1) (2011) 1–14.
[4] X.-S. Si, W. Wang, C.-H. Hu, M.-Y. Chen, D.-H. Zhou, A wiener-process-based degradation model with a recursive filter algorithm for remaining useful life estimation, Mech. Syst. Signal Process. 35 (1–2) (2013) 219–237.
[5] B. Saha, K. Goebel, J. Christophersen, Comparison of prognostic algorithms for estimating remaining useful life of batteries, Trans. Inst. Meas. Control 31 (3–4) (2009) 293–308.
[6] F.A. Niaki, M. Michel, L. Mears, State of health monitoring in machining: extended kalman filter for tool wear assessment in turning of in718 hard-to-machine alloy, J. Manuf. Process. 24 (2016) 361–369.
[7] L. Cui, X. Wang, Y. Xu, H. Jiang, J. Zhou, A novel switching unscented kalman filter method for remaining useful life prediction of rolling bearing, Measurement 135 (2019) 678–684.
[8] M.E. Orchard, G.J. Vachtsevanos, A particle-filtering approach for on-line fault diagnosis and failure prognosis, Trans. Inst. Meas. Control 31 (3–4) (2009) 221–246.
[9] P. Wang, R.X. Gao, Adaptive resampling-based particle filtering for tool life prediction, J. Manuf. Syst. 37 (2015) 528–534.
[10] H. Wang, C. Peng, Intelligent diagnosis method for rolling element bearing faults using possibility theory and neural network, Comput. Ind. Eng. 60 (4) (2011) 511–518.
[11] A.K. Mahamad, S. Saon, T. Hiyama, Predicting remaining useful life of rotating machinery based artificial neural network, Comput. Math. Appl. 60 (4) (2010) 1078–1087.
[12] Z. Tian, An artificial neural network method for remaining useful life prediction of equipment subject to condition monitoring, J. Intell. Manuf. 23 (2) (2012) 227–237.
[13] P. Welch, The use of fast fourier transform for the estimation of power spectra: a method based on time averaging over short, modified periodograms, IEEE Trans. Audio Electroacoust. 15 (2) (1967) 70–73.
[14] L. Liao, Discovering prognostic features using genetic programming in remaining useful life prediction, IEEE Trans. Industr. Electron. 61 (5) (2014) 2464–2472.
[15] R. Khelif, B. Chebel-Morello, S. Malinowski, E. Laajili, F. Fnaiech, N. Zerhouni, Direct remaining useful life estimation based on support vector regression, IEEE Trans. Ind. Electron. 64 (3) (2017) 2276–2285.
[16] Y. Yuan, H.-T. Zhang, Y. Wu, T. Zhu, H. Ding, Bayesian learning-based model-predictive vibration control for thin-walled workpiece machining processes, IEEE/ASME Trans. Mechatron. 22 (1) (2017) 509–520.
[17] M. Garnelo, D. Rosenbaum, C.J. Maddison, T. Ramalho, D. Saxton, M. Shanahan, Y.W. Teh, D.J. Rezende, S. Eslami, Conditional neural processes, Proceedings of the 35th International Conference on Machine Learning.
[18] M.E. Torres, M.A. Colominas, G. Schlotthauer, P. Flandrin, A complete ensemble empirical mode decomposition with adaptive noise, in: Acoustics, speech and signal processing (ICASSP), 2011 IEEE international conference on, IEEE, 2011, pp. 4144–4147.
[19] Z. Wu, N.E. Huang, Ensemble empirical mode decomposition: a noise-assisted data analysis method, Adv. Adaptive Data Anal. 1 (01) (2009) 1–41.
[20] N.E. Huang, Z. Shen, S.R. Long, M.C. Wu, H.H. Shih, Q. Zheng, N.-C. Yen, C.C. Tung, H.H. Liu, The empirical mode decomposition and the hilbert spectrum for nonlinear and non-stationary time series analysis, Proc.: Math. Phys. Eng. Sci. (1998) 903–995.
[21] J. Wu, C. Wu, S. Cao, S.W. Or, C. Deng, X. Shao, Degradation data-driven time-to-failure prognostics approach for rolling element bearings in electrical machines, IEEE Trans. Ind. Electron.
[22] M. Niu, Y. Wang, S. Sun, Y. Li, A novel hybrid decomposition-and-ensemble model based on CEEMD and GWO for short-term $PM_{2.5}$ concentration forecasting, Atmos. Environ. 134 (2016) 168–180.
[23] J.A. Suykens, J. Vandewalle, Least squares support vector machine classifiers, Neural Process. Lett. 9 (3) (1999) 293–300.
[24] T. Benkedjouh, K. Medjaher, N. Zerhouni, S. Rechak, Remaining useful life estimation based on nonlinear feature reduction and support vector regression, Eng. Appl. Artif. Intell. 26 (7) (2013) 1751–1760.
[25] C. Ordóñez, F.S. Lasheras, J. Roca-Pardiñas, F.J. de Cos Juez, A hybrid ARIMA–SVM model for the study of the remaining useful life of aircraft engines, J. Comput. Appl. Math. 346 (2019) 184–191.
[26] D. Liu, J. Zhou, D. Pan, Y. Peng, X. Peng, Lithium-ion battery remaining useful life estimation with an optimized relevance vector machine algorithm with incremental learning, Measurement 63 (2015) 143–151.
[27] Y. Miao, M. Gowayyed, F. Metze, Eesen: end-to-end speech recognition using deep rnn models and wfst-based decoding, in: Automatic Speech Recognition and Understanding (ASRU), 2015 IEEE Workshop on, IEEE, 2015, pp. 167–174.
[28] K. He, X. Zhang, S. Ren, J. Sun, Deep residual learning for image recognition, in: Proceedings of the IEEE Conference on Computer Vision and Pattern Recognition, 2016, pp. 770–778.
[29] F.A. Gers, J. Schmidhuber, F. Cummins, Learning to forget: continual prediction with LSTM, in: 9th International Conference on Artificial Neural Networks IET, 1999.
[30] G.E. Hinton, N. Srivastava, A. Krizhevsky, I. Sutskever, R.R. Salakhutdinov, Improving neural networks by preventing co-adaptation of feature detectors, arXiv preprint arXiv:1207.0580.
[31] D.P. Kingma, J.L. Ba, Adam: Amethod for stochastic optimization, in: Proc. 3rd Int. Conf. Learn. Representations, 2014.
[32] A. Saxena, K. Goebel, Turbofan engine degradation simulation data set, NASA Ames Prognostics Data Repository.
[33] F.O. Heimes, Recurrent neural networks for remaining useful life estimation, in: Prognostics and Health Management, 2008. PHM 2008. International Conference on, IEEE, 2008, pp. 1–6.
[34] A. Saxena, K. Goebel, D. Simon, N. Eklund, Damage propagation modeling for aircraft engine run-to-failure simulation, in: Prognostics and Health Management, 2008. PHM 2008. International Conference on, IEEE, 2008, pp. 1–9.
[35] Y. Yang, T. Pierce, J. Carbonell, A study of retrospective and on-line event detection, in: Proceedings of the 21st annual international ACM SIGIR conference on Research and development in information retrieval, ACM, 1998, pp. 28–36.
[36] P. Lim, C.K. Goh, K.C. Tan, P. Dutta, Multimodal degradation prognostics based on switching kalman filter ensemble, IEEE Trans. Neural Networks Learn. Syst. 28 (1) (2017) 136–148.
[37] G.S. Babu, P. Zhao, X.-L. Li, Deep convolutional neural network based regression approach for estimation of remaining useful life, in: International Conference on Database Systems for Advanced Applications, Springer, 2016, pp. 214–228.
[38] Y. Wu, M. Yuan, S. Dong, L. Lin, Y. Liu, Remaining useful life estimation of engineered systems using vanilla LSTM neural networks, Neurocomputing 275 (2018) 167–179.